\documentclass[number,12pt]{elsarticle}

\usepackage{amssymb}
\usepackage{amsmath}

\usepackage[hyphens]{url}
\usepackage{hyperref}
\usepackage{algorithm}
\usepackage{algpseudocode}
\usepackage{multirow}
\usepackage{subcaption}

\usepackage{graphicx}

\begin{document}

\begin{frontmatter}



\title{Incremental firmware update over-the-air for low-power IoT devices over LoRaWAN}


\author[label1]{Andrea De Simone} 
\ead{andrea.desimone@polito.it}


\author[label1]{Giovanna Turvani} 


\author[label1]{Fabrizio Riente} 

\affiliation[label1]{organization={Department of
Electronics and Telecommunication},
            addressline={Politecnico di Torino}, 
            city={Turin},
            postcode={10129}, 
            country={Italy}}

\begin{abstract}
Efficiently supporting remote firmware updates in Internet of Things (IoT) devices remains a significant challenge due to the limitations of many IoT communication protocols, which often make it impractical to transmit full images over low-bandwidth links such as LoRaWAN. Techniques such as firmware partitioning have been introduced to mitigate this issue, but they frequently fall short, especially in battery-powered systems where time and energy constraints are critical. As a result, physical maintenance interventions are still commonly required, which is costly and inconvenient in large-scale deployments. In this work, we present a lightweight and innovative method that addresses this challenge by generating highly compact delta patches, enabling firmware reconstruction directly on the device. Our algorithm is specifically optimized for low-power devices, minimizing both memory usage and computational overhead. 
Evaluated on eight firmware pairs for an STM32WL-based  IoT node, \textit{bpatch} shrinks the payload by up to 680 × for near-unchanged updates and by 9-18 × for typical minor revisions, significantly reducing the data volume needed for updates while maintaining performance comparable to more complex alternatives.
Experimental evaluations confirm that our method yields substantial time and energy savings, making it particularly well-suited for battery-powered IoT nodes. \textit{bpatch} is released as open-source and, although demonstrated on LoRaWAN, the approach is flexible and can be adapted to other IoT communication technologies.
\end{abstract}




\end{frontmatter}



\section{Introduction}
In the Internet of Things (IoT) scenario, system reliability and durability are essential factors to ensure efficient and prolonged operation. Throughout the lifecycle of IoT devices, maintenance activities are often necessary, making device maintainability a key consideration already at the design stage. One of the distinctive attributes of IoT devices is their autonomy and self-sufficiency, especially since they are often deployed in remote or hard-to-reach locations, or in large numbers, thus making individual manual maintenance impractical.
A common and critical maintenance task in IoT systems is firmware updating. The motivations behind firmware updates are manifold, including bug fixes, performance enhancements, or adaptation to new operational requirements. Among the various communication technologies used in IoT, the LoRaWAN protocol is widely recognized as a standard for low-power, long-range data transmission \cite{ref17}. 
The Things Network, one of the most adopted LoRaWAN-based infrastructures, supports remote firmware updates via the Firmware Update Over-the-Air (FUOTA) protocol package \cite{ref1}. LoRaWAN's main strengths lie in its energy efficiency and extensive communication range-up to tens of kilometers \cite{ref18}. However, its limited bandwidth poses a significant challenge for the transmission of large payloads, such as complete firmware images.
Traditional FUOTA methods address this limitation by fragmenting the firmware into smaller packets and transmitting them sequentially. This procedure, however, is both time- and energy-consuming. Numerous studies have attempted to improve the remote firmware update process \cite{ref8}. Analysis of power consumption in LoRaWAN-enabled microcontroller units (MCUs) has shown that the radio transceiver is the dominant energy consumer during FUOTA \cite{ref7, ref9}. Class C devices are often used due to the high data throughput required, even though they consume more energy than the more efficient Class A devices, as their radios remain constantly active to receive packets \cite{ref5}. Another relevant factor is the Spreading Factor (SF), which directly impacts the power and time required for data reception \cite{ref11}.

To reduce energy consumption during FUOTA, two main strategies are explored in the literature. The first is the use of multicast transmissions, which allow simultaneous broadcasting of firmware fragments to multiple devices. Optimization techniques such as grouping nodes by their SF have been proposed to maximize efficiency \cite{ref13, ref14}.

The second promising technique consists of reducing the transmitted data volume. Approaches include modular code deployment with dynamic linking \cite{ref12}, or structured code and regular function addressing \cite{ref6, ref10,ref15}. 
These methods, while reducing the transmission size and sometimes eliminating the need for rebooting, are often hardware-dependent, limiting their general applicability.
A more flexible alternative is to compute only the differences \textit{delta} between old and new firmware version, transmitting only essential information for reconstructing the updated firmware.

The base commands in delta files are \textit{COPY} to reuse existing parts from old firmware and \textit{ADD} to insert new bytes, as exemplified in \cite{ref2, ref3,ref4, ref13}. Each paper explores enhancements on the base differential algorithm. For instance, \cite{ref2} uses an XOR delta algorithm to reduce the complexity of the algorithm, whereas \cite{ref3} introduces a \textit{REPAIR} command for copying large segments with minor differences. Additionally, variations of opcodes of the same command with different lengths are available to optimize the patch size.
The study in \cite{ref4} optimizes the research of common segments and the sequence of the transmission, while \cite{ref13} employs a Suffix Array-based delta algorithm, optimizing the segments research, otherwise they achieve a limited compression in comparison to the other references. 
Overall, existing methods for compacting delta files typically rely on complex encoding techniques for opcodes or exploit specific characteristics of binary instructions to achieve good compression results.

In this paper, we propose a novel approach to perform incremental updates, named \textit{bpatch}, which is open-accessible at \cite{zenodo-bpath}. 
The goal is to design a hardware-independent and lightweight solution suitable for integration into low-resource IoT devices. Unlike other techniques that rely on compression algorithms or firmware structure constraints, \textit{bpatch} only uses the fundamental \textit{COPY} and \textit{ADD} instructions, encoded at bit-level for minimal overhead. This simplicity enables efficient delta generation and firmware reconstruction without sacrificing compression performance.
Importantly, the focus of this work is on the application of the delta patch, not on the generation process or the security/authenticity aspects of the update procedure. Issues such as secure patch delivery, cryptographic verification, or delta algorithm design fall outside the scope of this paper and are not addressed here.
Results indicate that \textit{bpatch} achieves compression sizes comparable to more complex algorithms, which are not suitable for integration into hardware-constrained MCUs. Consequently, this method enables effective FUOTA on battery-operated IoT devices, with limited resources, by significantly reducing energy usage associated with firmware transmission and the complexity of reconstruction.

To validate the effectiveness of \textit{bpatch}, we benchmarked it against popular open-source delta algorithms, including \textit{bsdiff} \cite{ref_bsdiff}, \textit{VCDIFF} \cite{ref_vcdiff} and \textit{HDiffPatch} \cite{ref_hdifpatch}. These algorithms are generally employed to compress Android Package Kit (APK) files for mobile applications and are not suitable for low-power IoT devices.
Furthermore, we provide an actual implementation of \textit{bpatch} and evaluate its energy savings compared to traditional FUOTA methods in a real-world scenario.
The primary contributions of the manuscript include:
\begin{itemize}
    \item the design of an innovative methodology for efficiently encoding an edit script without relying on additional compression algorithms;
    \item the creation of a hardware-independent reconstruction algorithm that requires minimal MCU hardware resources;
    \item the proposed algorithm was benchmarked using 173 delta scripts, derived from real firmware projects, across three different architectures to assess its compression capability;
    \item the experimental deployment of \textit{bpatch} in an ultra-low-power IoT node, which includes the STM32WL55JC MCU, demonstrating reduced update duration and energy usage compared to standard updates;
    \item the public release of the \textit{bpatch} algorithm on Zenodo \cite{zenodo-bpath} with a ready-to-use application to generate patches and reconstruct the final firmware.  
\end{itemize}

The rest of this article is organized as follows. 
Section~\ref{sec_arch} gives an overview of a real architecture where \textit{bpatch} is implemented.
Section~\ref{sec_alg} provides a detailed description of the algorithm that we developed.
In Section~\ref{sec_comp} we compare the performance of \textit{bpatch} with other open-source solutions.
Section~\ref{sec_en} analyzes the energy consumption of FUOTA with and without \textit{bpatch} and finally, Section~\ref{sec_concl} draws conclusions. 

\section{System architecture and FUOTA integration with \textit{bpatch}} \label{sec_arch}

\begin{figure}[htbp]
\centering
\begin{subfigure}[c]{0.48\textwidth}
    \centering
    \includegraphics[width=\linewidth]{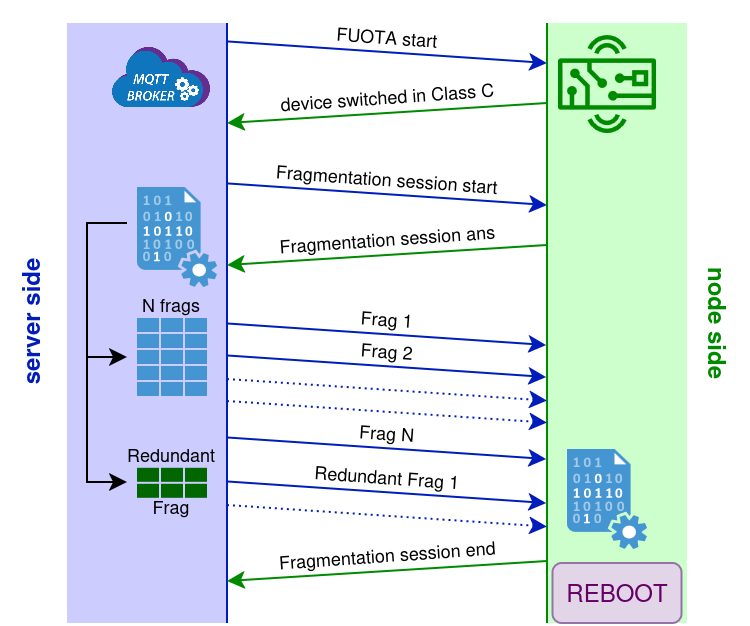}
    \caption{}
    \label{fig:fuota_a}
\end{subfigure}
\hfill
\begin{subfigure}[c]{0.48\textwidth}
    \centering
    \includegraphics[width=\linewidth]{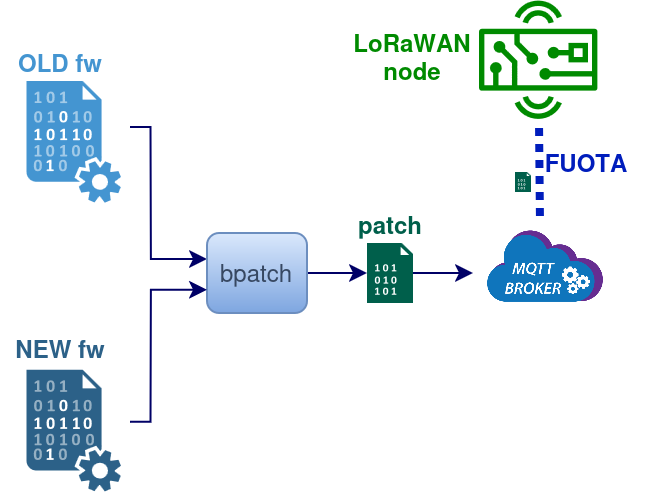}
    \caption{}
    \label{fig:fuota_b}
\end{subfigure}
\caption{Overview of the FUOTA protocol along with \textit{bpatch} integration.}
\label{fig_fuota}
\end{figure}

Firmware updates in low-power IoT devices introduce a critical trade-off between functionality and energy efficiency. Traditional FUOTA mechanisms transmit entire firmware images, requiring prolonged radio activity that directly contradicts the ultra-low-power operational model of typical IoT applications. This section presents the real-world implementation of a FUOTA-enabled architecture where the proposed \textit{bpatch} algorithm has been integrated to address these challenges.
To provide a concrete context, we consider an IoT system for monitoring the health status of bee hives \cite{ref19}. In this application, firmware updates are rare but unavoidable, and energy preservation is paramount.
The MCU used belongs to the STM32WL family, designed for ultra-low power operation and native LoRaWAN integration. Based on the guideline provided by ST Microelectronics \cite{ref_an5554}, we successfully deployed a standard FUOTA process for support complete firmware replacement. The LoRaWAN node operates at 868 MHz with SF 9, providing a maximum payload size of 115 bytes per message. Due to Fair Use Policy limitations on duty cycles, a maximum of 530 downlink messages per hour is allowed, approximately one message every 7 seconds. 
The transmission of packets does not require acknowledgment. Lost fragments are recovered using redundant fragments transmitted on-demand: after sending all initial fragments, additional redundancy fragments continue to be transmitted until the node confirms successful firmware assembly.
The server-side management is performed via a Message Queuing Telemetry Transport (MQTT) Broker, which is responsible for sending commands, including fragment transmissions, and receiving node responses. A summary of the communication protocol is illustrated in Fig.~\ref{fig_fuota}(a), the server side is represented in blue, and the blue arrows indicate the downlinks sent to the node, while the node is represented in green, and the green arrows indicate the uplinks.
Upon receiving the complete firmware code, a Firmware Update Agent (FUA) initiates a reboot using the new firmware image. After performing an integrity check on the new binary image using a cryptographic key, the new firmware replaces the previous one, finalizing the update process. This procedure is thoroughly described in \cite{ref_an5056}. 
Given a target application firmware size of approximately 100 KB, more than 900 fragments are necessary (considering 112 bytes per fragment, because 3 bytes are used by the LoRaWAN protocol), resulting in a transmission duration that exceeds 100 minutes. The radio must remain continuously active during this period, significantly increasing energy consumption. Such prolonged activity conflicts with the system’s intended operational design, which aims to stay in low-power mode most of the time, activating only for some seconds every few hours. Consequently, despite the infrequent nature of firmware updates, their associated energy costs remain excessively high.

\begin{figure}[t]
\centering
\includegraphics[width=0.8\linewidth]{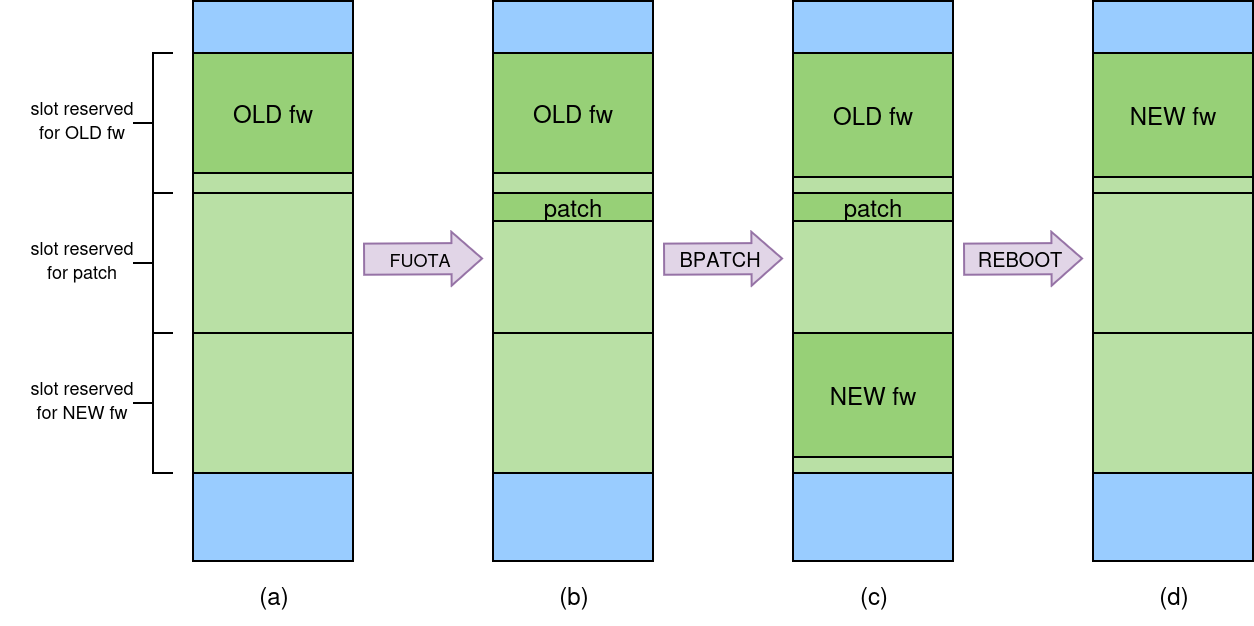}
\caption{Footprint of the MCU's flash.}
\label{fig_memory}
\end{figure}

To mitigate the inefficiency of full-image transmission, we integrated the proposed \textit{bpatch} algorithm into the same architecture. By adopting the \textit{bpatch} algorithm, the delta file transmitted is several times smaller than the entire firmware. Moreover, the existing communication protocol remains unaltered, with the only additional step being the patching operation to reconstruct the new firmware before rebooting. An overview of the modified procedure is provided in Fig.~\ref{fig_fuota}(b), while Fig.~\ref{fig_memory} provides details regarding the memory footprint of the improved system. Specifically, Fig.~\ref{fig_memory}(a) illustrates the initial state of memory, Fig.~\ref{fig_memory}(b) represents the memory state upon receiving the patch, Fig.~\ref{fig_memory}(c) depicts the system state after reconstructing the updated firmware image, and Fig.~\ref{fig_memory}(d) shows the final state where the new firmware replaces the previous version.
As detailed in Section~\ref{sec_alg}, \textit{bpatch} introduces no complex computational requirements. This design ensures that the reconstruction process remains feasible on hardware-constrained MCUs without compromising the low-power objectives of the IoT node.


\section{Description of \textit{bpatch} algorithm}  \label{sec_alg}

The desired characteristic of \textit{bpatch} is to produce an extremely compact delta file that contains only simple instructions to construct the new file from the older one. It should be platform-independent and must be specific for binary files. In the following section, some considerations are exposed to illustrate the main idea of \textit{bpatch}. Here, as a case study, it is applied to LoRaWAN, but the approach could be used to other protocols. 

While binary firmware images inherently depend on specific hardware architectures, we notice that even small code modifications tend to generate unpredictable and widespread differences. As investigated in \cite{ref3}, inserting or deleting functions or global variables can shift physical addresses, resulting in minor byte-level changes across the entire firmware image. Consequently, typical firmware modifications, such as adding a few functions or minor code adjustments, can produce small and pervasive differences in the entire firmware. 

Considering this factor along with the requirements of \textit{bpatch}, the proposed encoding strategy includes the following features:
\begin{itemize}
    \item Only the two fundamental commands, \textit{COPY} and \textit{ADD}, are utilized. 
    \item Byte positions in the delta file are expressed as incremental values rather than absolute positions, to reduce the range of numbers.
    \item Bit lengths allocated for command fields are dynamically customized rather than fixed.
    \item Each field is preceded by an indication of its bit length.
    \item No bits are wasted in indicating the command type (\textit{COPY} or \textit{ADD}), alternation maintaining the consistency.
\end{itemize}

The first phase of the algorithm consists of finding the difference between the old and new versions of the firmware. For this purpose, the UNIX diff program is exploited, and the binary files are converted into text files with one byte per row before applying the program. This choice is adopted because UNIX diff \cite{ref_diffutils} is based on the Myers algorithm. This algorithm aims to find the shortest number of instructions that transform the first text file into the second one \cite{ref20, ref21}. The instructions can be easily decomposed into \textit{COPY} and \textit{ADD} opcodes, which follow previously listed rules.

\begin{figure}[h]
\centering
\includegraphics[width=0.6\linewidth]{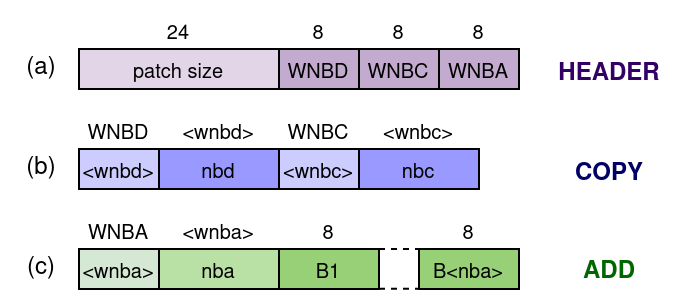}
\caption{Opcodes employed in \textit{bpatch}: (a) header, (b) \textit{COPY} and (c) \textit{ADD}.}
\label{fig_opcode}
\end{figure}

The innovation of this encoding lies in the optimal compression achieved by customizing bit allocation for opcode fields, thereby avoiding unnecessary bit wastage. The opcode formats are illustrated in Fig.~\ref{fig_opcode}, where each rectangle represents a field value, and the corresponding bit length is indicated above each field.
Specifically, the \textit{COPY} opcode, depicted in Fig.~\ref{fig_opcode}(b), includes two fields: \textit{nbd} (number of bytes to discard in the original firmware) and \textit{nbc} (number of bytes to copy from the original firmware). Conversely, the \textit{ADD} opcode, described in Fig.~\ref{fig_opcode}(c), includes a single field \textit{nba}, representing the number of new bytes to insert, followed by the corresponding byte values. 
Bit lengths for each field are dynamically adjusted based on the values they must represent, and the bit length itself is indicated explicitly before each field. Consequently, smaller numerical values, very common in this case, consume fewer bits rather than fixed byte-lengths. 
The bit length of these pre-fields is fixed by constants (WNBD, WNBC and WNBA) established before generating the delta file, depending on the maximum of their values. These constants are contained in the patch inside a header, Fig.~\ref{fig_opcode}(a), along with the patch size expressed in \textit{bits}.

\begin{algorithm}
\caption{bpatch reconstruction}
\begin{algorithmic}
\State patch\_size $\gets$ \textsc{read}(24) \Comment{start header} 
\State WNBD $\gets$ \textsc{read}(8) \Comment{read from patch 8 bits}
\State WNBC $\gets$ \textsc{read}(8)
\State WNBA $\gets$ \textsc{read}(8)
\State i $\gets$ 0
\State \While{i $<$ patch\_size} 
    \State wnbd $\gets$ \textsc{read}(WNBD) \Comment{start COPY opcode}
    \State nbd $\gets$ \textsc{read}(wnbd)
    \State \textsc{jump}(nbd) \Comment{skip \textit{nbd} bytes from old}
    \State i $\gets$ i + WNBD + wnbd \Comment{update patch index}
    \State wnbc $\gets$ \textsc{read}(WNBC)
    \State nbc $\gets$ \textsc{read}(wnbc)
    \State \textsc{copy}(nbc) \Comment{copy \textit{nbc} bytes from old to new}
    \State i $\gets$ i + WNBC + wnbc
    \State wnba $\gets$ \textsc{read}(WNBA) \Comment{start ADD opcode}
    \State nba $\gets$ \textsc{read}(wnba)
    \State \textsc{insert}(nba) \Comment{copy \textit{nba} bytes from patch to new}
    \State i $\gets$ i + WNBA + wnba + 8$\cdot$nba
\EndWhile

\end{algorithmic}
\label{alg_bpatch}
\end{algorithm}

An Application Program Interface (API) is implemented in Python to handle \textit{bpatch}. Firmware reconstruction is based on a dedicated C function, which loads the patch and original firmware into memory buffers and assembles the updated firmware according to patch instructions. This function can be integrated into an MCU to perform the FUOTA.
The Python API and the C reconstruction software is openly available on GitHub \url{https://github.com/vlsi-nanocomputing/bpatch} and archived on Zenodo \cite{zenodo-bpath}. A concise summary of the reconstruction algorithm is provided in Algorithm~\ref{alg_bpatch}. 
Initially, the patch size and the lengths of fixed fields are read from the header. Subsequently, an alternating sequence of \textit{COPY} and \textit{ADD} commands specifies the bytes to copy from the existing firmware version and the new bytes to insert, respectively. As previously noted, fields are not byte-aligned, and therefore reading operations are performed at the bit level.

\section{Comparison on delta algorithms}  \label{sec_comp}

The compression performance of \textit{bpatch} was assessed and compared with three differential open-source algorithms, \textit{bsdiff}, \textit{vcdiff} and \textit{HDiffPatch}, which are widely used for updating mobile applications \cite{ref16} thanks to their ability to produce very compact patch files.
However, it should be noted that smartphones and tablets offer orders of magnitude more memory, processing power, and energy than ultra-low-power MCUs typically found in IoT nodes. An algorithmic complexity that is acceptable in the mobile domain can become prohibitive on constrained devices. A brief overview of each algorithm is provided below:
\begin{itemize}
    \item \textbf{bsdiff}: Utilizes a Suffix Array approach to identify differences, and it can recognize not only identical segments, but also partially similar segments by transmitting just the differing bytes.
    This algorithm typically generates large patch files filled with many zeros and repetitive sequences, which are subsequently compressed using \textit{bzip2}, allowing for significantly reduced final patch sizes.
    
    \item \textbf{VCDIFF}: Employs a straightforward delta encoding approach based on byte-level \textit{COPY} and \textit{ADD} opcodes, with minimal complications. While the algorithm itself does not include compression, external compression is commonly applied to reduce the final patch size.
    
    \item \textbf{HDiffPatch}: Similar to \textit{bsdiff} in principle, \textit{HDiffPatch} provides considerably improved performance in terms of both computational efficiency and reduced patch size. As with \textit{VCDIFF}, external compression can further decrease patch size but is not inherently integrated into the algorithm.
\end{itemize}

To assess performance, three Git repositories tracking firmware evolution for the selected MCU in three different real-world applications were analyzed. For consistency, compression strategies were disabled during testing, specifically, a variant of \textit{bsdiff} without the \textit{bzip2} compression stage was employed.

\begin{figure}
  \centering
  \subfloat[]{\includegraphics[height=5cm]{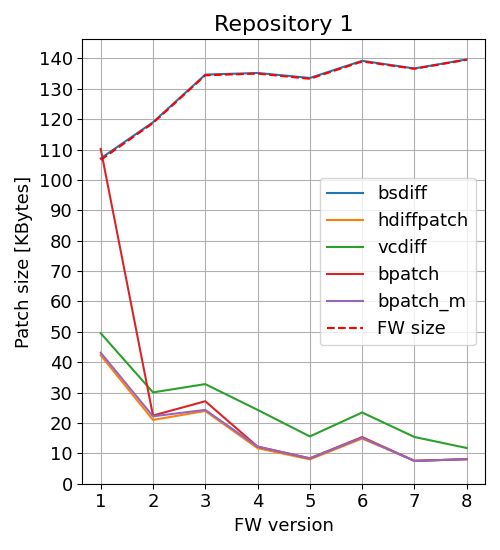}}
  \hfill
  \subfloat[]{\includegraphics[height=5cm]{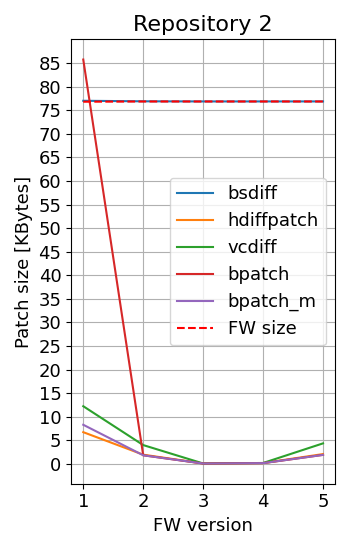}}
  \hfill
  \subfloat[]{\includegraphics[height=5cm]{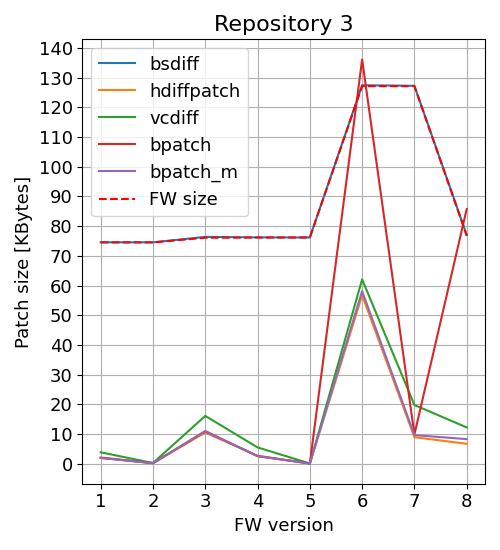}}
  \caption{Comparison of patch sizes produced by the five algorithms for the three Git repositories (a), (b), and (c).}
  \label{fig_comparison}
\end{figure}

Fig.~\ref{fig_comparison} compares patch sizes generated by each algorithm. Two variants of the \textit{bpatch} are presented: the standard \textit{"bpatch"}, which employs the basic UNIX diff command, and the enhanced \textit{"bpatch\_m"}, which incorporates the "minimal" option (available using option "-d"). This option applies an optimized variant of the diff algorithm to further compress the set of changes, potentially at the cost of increased processing time. Attempts to use alternative differential algorithms, such as the pure Myers algorithm described in \cite{ref20} (UNIX diff strategies propose a slight variation of Myers algorithm), did not yield significant improvements over \textit{bpatch\_m}, thus, they were excluded from the analysis.

From the three graphs, \textit{HDiffPatch} and \textit{bpatch\_m} consistently generate the smallest patches, whereas \textit{VCDIFF} produces slightly larger delta files, differing by only a few kilobytes. Conversely, \textit{bsdiff} generates patches nearly equivalent in size to the original firmware, reflecting its dependence on external compression (\textit{bzip2}) to achieve reduced patch sizes. The standard \textit{bpatch}, utilizing the basic UNIX diff command, generally matches the performance of its enhanced counterpart, \textit{bpatch\_m}, except in certain cases where patch sizes are equal to or exceed the original firmware size.

To further analyze the hardware independence and general performance of \textit{bpatch}, an additional evaluation was conducted using an open-source firmware repository containing incremental firmware versions. The ArduPilot website \cite{ref_ardupilot} provides numerous binary images used for drone boards. This source was chosen due to its variety of small-size binaries (under 2 MB), multiple supported boards, and numerous sequential firmware versions. Specifically, five project were considered: \textit{Plane}, \textit{Copter}, \textit{Rover}, \textit{Sub} and \textit{Antenna Tracker}, inside them two folders were selected: \textit{ACNS-CM4Pilot}, based on STM32H7 microcontrollers, and \textit{navio}, based on Raspberry Pi boards. In total, 173 patches were generated from firmware images obtained from the three previous repositories and these two folders. 
Tests on larger binaries were not conducted for two main reasons. First, the proposed solution is specifically designed for hardware-constrained MCUs, which typically do not handle large firmware images. Second, the UNIX diff, based on the Myers algorithm, is not optimized for large files with extensive differences.

To provide clear insights, three firmware update scenarios were identified:
\begin{itemize}
    \item \textbf{Major updates (MJ)}: Significant code refactoring, characterized by substantial code modifications and high byte-count variations.
    \item \textbf{Minor updates (MN)}: Only limited changes such as bug fixes, feature enhancements, or minor code adjustments. This scenario represents the primary target application for \textit{bpatch}.
    \item  \textbf{No updates/constant updates (NU)}: Situations where firmware versions remain essentially unchanged (e.g., documentation updates) or contain minimal byte differences (e.g. macro or constant modifications). While remote updates may be unnecessary in these cases, testing was conducted for completeness.
\end{itemize}

Patch sizes generated by \textit{HDiffPatch} were used to categorize these scenarios: patches smaller than 0.5\% of the total firmware size were classified as NU, whereas patches exceeding 20\% were classified as MJ. \textit{HDiffPatch} was selected as the reference due to its superior compression among the analyzed open-source tools.

Table~\ref{tab_compression} reports the \emph{compression factor} defined as
$\text{image size} / \text{patch size}$ for every algorithm and platform.
\textit{bsdiff} was excluded from this comparison, as previously noted, it relies on an external \textit{bzip2} pass for generating small patch files. The column labeled "STM32WL" corresponds to the Git repositories illustrated in Fig.~\ref{fig_comparison}, while the remaining two columns pertain to binaries from ArduPilot.

Both \textit{bpatch} variants dominate the NU scenario.  
On STM32WL the factor reaches $\sim$\,680×, versus 561× for \textit{HDiffPatch} and 591× for \textit{VCDIFF}.  
The gap widens on the larger STM32H7 and Raspberry Pi images, where \textit{bpatch} attains 33k-39k× while the best competitor stays below 7k×.  
Hence, when only a handful of bytes change, as is common in IoT hot-fixes, \textit{bpatch} delivers an order-of-magnitude better size reduction.

In the case of the minor updates (MN) scenario, \textit{bpatch\_m} equals or slightly surpasses \textit{HDiffPatch} on every benchmark (18.0× vs.\ 17.8× on STM32WL, 15.1× vs.\ 15.5× on STM32H7,  9.37× vs.\ 10.13× on Raspberry Pi). The standard \textit{bpatch} follows within a few percent, while \textit{VCDIFF} lags.

When more than 20\% of the firmware changes, the patch overhead becomes appreciable. The \textit{bpatch} variant falls below unity on STM32WL (the patch is slightly larger than the new image), but the optimized \textit{bpatch\_m} recovers a 2.3-3.3× factor—essentially matching \textit{HDiffPatch} (2.4-4.0×) and outperforming \textit{VCDIFF} on two out of three platforms.

In summary, \textit{bpatch\_m} offers compression comparable to (and often better than) state-of-the-art algorithms across all scenarios, while standard \textit{bpatch} excels whenever the modification is small.  
Conversely, both \textit{HDiffPatch} and \textit{VCDIFF} demand a filesystem and several resources to decompress patches, resources that are unavailable on many MCUs. By contrast, the constant-footprint reconstruction routine of \textit{bpatch} runs entirely in place and is therefore immediately deployable on ultra-low-power IoT nodes.

\begin{table}[h]
\centering
\begin{tabular}{ c c c c c c c c c c}
\hline

& \multicolumn{3}{c}{STM32WL} & \multicolumn{3}{c}{STM32H7} & \multicolumn{3}{c}{Raspberry Pi}\\
 & MJ & MN & NU & MJ & MN & NU & MJ & MN & NU \\
\hline

bpatch & 0.951 & 16.9 & 676 & 1.46 & 15.0 & 33844 & 1.47 & 9.30 & 39067\\
bpatch\_m & 2.33 & 18.0 & 676 & 3.31 & 15.1 & 33844 & 2.34 & 9.37 & 39067\\
hdiffpatch & 2.38 & 17.8 & 561 & 4.04 & 15.5 & 6793 & 2.64 & 10.13 & 6955\\
vcdiff & 2.1 & 9.5 & 591 & 2.06 & 4.3 & 14396 & 1.64 & 3.43 & 13568\\

\hline

\end{tabular}
\caption{Average compression for the four algorithms for three different architectures.} \label{tab_compression}
\end{table}

\section{Energy measurement and update time evaluation of \textit{bpatch}}  \label{sec_en}

This section quantifies the energy and time savings obtained by \textit{bpatch} in the beehive monitoring system introduced in Section~\ref{sec_arch}.  
Competing delta tools are not considered here because, as discussed earlier, their reconstruction routines require a filesystem and RAM/CPU resources that are unavailable on the target MCU; a fair on-device comparison is therefore impossible. The test system is built around an STM32WL55JC microcontroller, equipped with 256 kB of on-chip flash and 64 kB of SRAM, and an external 8 Mbit MX25L8006EM1I-12G flash device.

Typically, the IoT node remains in sleep mode when not actively performing measurements, during which the MCU and peripherals are powered down, leaving only the Real-Time Clock (RTC) operational to wake up the MCU periodically. However, during FUOTA operations, the radio module must remain active to receive firmware fragments. LoRaWAN specifies three operational classes defining device behavior for downlink reception, which reflect the operating mode of the radio, the classes \cite{ref_lorawan_classes} are:  
\begin{itemize}
    \item \textbf{Class A}: After each uplink transmission, the device opens two brief receiving windows, thus limiting radio activation.
    \item \textbf{Class B}: Extends Class A functionality by periodically opening additional receiving windows (ping slots) synchronized via beacons transmitted by the server side
    \item \textbf{Class C}: The radio remains continuously active, always ready to receive downlink transmissions.
\end{itemize}
Under normal operating conditions, the device operates in Class A mode to minimize energy by deactivating the radio, but switches to Class C for the duration of a FUOTA session. We quantified the impact of transitioning between these classes and the FUOTA process by measuring the current consumption in sleep mode for the two classes.
Measurements were taken using a Tektronix DMM7510 digital multimeter, with the IoT device powered at 3.6 V, corresponding to the nominal battery voltage. Current consumption was recorded at sample rates between 1 kHz and 100 kHz, depending on measurement duration.

\begin{table} [h]
\centering
\begin{tabular}{c c c c}
\hline
& Avg current & Std current & Sample Rate \\
\hline
Class A     & 1.75 µA & 0.37 µA & 1k/s \\
Class C     & 6459.8 µA & 7.5 µA & 1k/s \\
\hline
\end{tabular}
\caption{Current measurements for Class A and C in low-power mode.} \label{tab_class_current}
\end{table}

Table~\ref{tab_class_current} summarizes the current consumption comparison between the two operational classes over a 25-minute duration, indicating average (Avg) and standard deviation (Std). Results clearly demonstrate that Class C consumes significantly more energy, over three orders of magnitude greater than Class A. Consequently, reducing the FUOTA duration through fewer transmitted fragments directly translates into substantial battery energy savings.

\begin{figure}[h]
\centering
\includegraphics[width=0.7\linewidth]{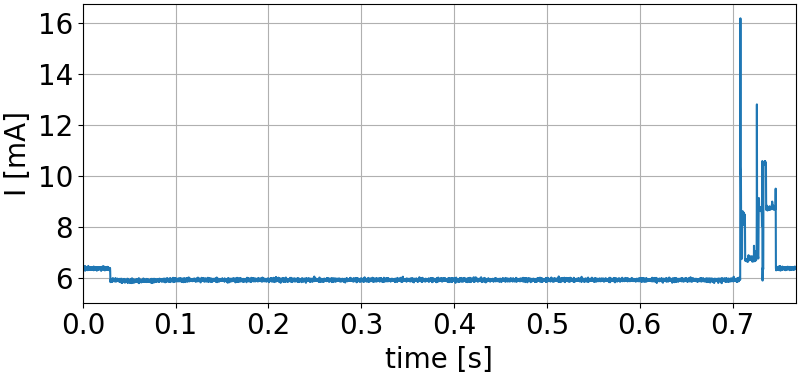}
\caption{Current measurements during the transmission of one fragment.}
\label{fig_frag}
\end{figure}

A subsequent measurement aimed at estimating the current consumption per firmware fragment received was performed, thus establishing a baseline for the minimum energy required during FUOTA. We measured current consumption during the reception of 135 fragments and derived estimates for time and energy. Fig.~\ref{fig_frag} provides a representative measurement for receiving one fragment. 
Initially, the device operates in Class C mode with the radio actively listening, consuming approximately 6.5 mA. Subsequently, the fragment is received, during which the current consumption slightly decreases to about 5.9 mA. Afterwards, the fragment is written into the MCU's memory, and finally, the node returns to Class C listening mode.
Detailed results are reported in Table~\ref{tab_frag_current}.

\begin{table}[h]
\centering
\begin{tabular}{c c c c}
\hline
& Avg & Std & Sample Rate \\
\hline
Current     & 6060.58 µA & 12 µA & \multirow{2}{*}{8k/s} \\
Time        & 18.8 ms & 3.3 ms & \\
\hline
\end{tabular}
\caption{Fragment current and time measurements.} \label{tab_frag_current}
\end{table}

Using these measurements, we estimated the total energy required for transmitting the full firmware update versus using incremental updates combined with \textit{bpatch} (with the "minimal" option enabled). Energy consumption is expressed in $mAh$ to intuitively assess battery impact. 
For the measurements are take as reference the FW versions in Repository 1. 
Table~\ref{tab_energy} presents the comparative analysis, highlighting substantial improvements in time and energy when employing incremental updates. Specifically, the table is structured as follows: the three \textit{Full-image update} and \textit{Incremental update} columns list, in order, the fragment count, the transmission time, and the energy consumed when the update is performed with \textit{bpatch}. The last two columns indicate the differences in these metrics between the two approaches.
The benefits of incremental update are undeniable, both in terms of time and energy, indeed reducing transmission duration by up to 18 times, achieving proportional energy savings.

\begin{table}[h]
\centering
\begin{tabular}{c  c  c c  c  c c c}
\hline
   \multicolumn{3}{c}{Full-image update}& \multicolumn{3}{c}{Incremental update} & \multicolumn{2}{c}{Reduction} \\
\multirow{2}{*}{Frags} & Time  & Energy & \multirow{2}{*}{Frags} & Time & Energy & Time   & Energy \\
 & [min] & [mAh]  &  & [min]& [mAh]  & [ratio]& [mAh] \\
\hline

975 & 114 & 12.17 & 395 & 46 & 4.93 & 2.47 & 7.24 \\
1086 & 127 & 13.55 & 203 & 24 & 2.53 & 5.35 & 11.02 \\
1230 & 144 & 15.35 & 223 & 26 & 2.78 & 5.52 & 12.57 \\
1235 & 144 & 15.41 & 112 & 13 & 1.40 & 11.03 & 14.02 \\
1219 & 142 & 15.21 & 77 & 9 & 0.96 & 15.83 & 14.26 \\
1271 & 148 & 15.86 & 139 & 16 & 1.73 & 9.14 & 14.13 \\
1249 & 146 & 15.59 & 69 & 8 & 0.86 & 18.10 & 14.73 \\
1276 & 149 & 15.93 & 74 & 9 & 0.92 & 17.24 & 15.01 \\

 \hline

\end{tabular}
\caption{Energy and time reduction for update using \textit{bpatch} on Repository 1.} \label{tab_energy}
\end{table}

It is important to note that these energy estimations consider only the fragment transmission phase, which represents the most prolonged and energy-intensive portion of the FUOTA process. They exclude the initial protocol setup and final device reboot phases, thus representing a conservative lower bound of total FUOTA energy consumption. 

In case of incremental update, an additional computational overhead associated with firmware reconstruction must be taken in account. For this reason, we measured it on a selected set of firmware patches, labeled as MN.
Results are presented in Table~\ref{tab_bpatch}, the columns describe the firmware version selected, the average reconstruction time, and the associated energy consumption, averaged over five separate measurements.
The reconstruction time is primarily determined by I/O operations, specifically reading the old firmware and the patch file, and subsequently writing the new firmware image.
The energy required exhibits slight variations related to the new firmware size, it is considerably smaller, approximately three orders of magnitude lower than the energy consumption during the fragment transmission phase. Therefore, this additional reconstruction overhead can be considered negligible relative to the overall energy savings previously discussed.

\begin{table} [h]
\centering
\begin{tabular}{ c  c  c  c  c }
\hline
Patch  & New FW & Average  & Average & Estimated \\
Size [KB]   & Size [KB]         & Current [mA]  & Time [ms]    & Energy [µAh]    \\
\hline
22.44 & 118.72 & 12.14 & 4635.5 & 15.64 \\
12.21 & 135.00 & 11.92 & 5028.2 & 16.65 \\
8.35 & 133.28 & 11.89 & 4990.7 & 16.48 \\
15.37 & 138.98 & 11.86 & 5117.1 & 16.85 \\
7.50 & 136.58 & 11.86 & 5065.7 & 16.69 \\
 \hline

\end{tabular}
\caption{Energy and time estimations for \textit{bpatch}.} \label{tab_bpatch}
\end{table}

\section{Conclusions}  \label{sec_concl}

This study presents a simple yet effective method for remotely updating IoT devices, achieving high compression of firmware images while ensuring minimal computational requirements during firmware reconstruction, thus making it feasible even for hardware-constrained microcontrollers and energy-constrained systems. 
The proposed solution is independent of the firmware binary structure and the architecture of the MCU.
The differential algorithm leverages the robust and well-established UNIX diff software, enhanced by a carefully optimized representation of the edit script for effective compression. The entire software is open-source and available for testing and custom implementation.
Benchmarking \textit{bpatch} against other open-source patch-generation algorithms demonstrates its ability to compress firmware sizes. A real-world evaluation further confirms that incremental updates employing \textit{bpatch} substantially reduce both transmission time and energy consumption compared to full firmware updates. Specifically, minor updates achieve a compression ratio ranging from 9 to 18 times, resulting in a proportional reduction in energy consumption and transmission time.
In conclusion, \textit{bpatch} represents an effective and energy-aware solution for FUOTA, particularly suitable for battery-powered IoT applications, significantly reducing the impact on battery lifespan.

Future work will explore enhancements to the edit script structure and the potential introduction of additional or specialized opcodes to achieve even better compression. A key challenge will be maintaining simplicity in both structure and firmware reconstruction processes. 
Finally, while data compression programs commonly utilized by other differential algorithms have not been evaluated in this study, due to the primary objective of minimizing computational overhead at the edge, their integration could be beneficial for more capable hardware platforms to further reduce patch sizes and optimize data transmission during FUOTA.

\bibliographystyle{elsarticle-num}
\bibliography{bibliography}





\end{document}